# Latency-Sensitive 5G RAN Slicing for Deterministic Aperiodic Traffic in Smart Manufacturing

M. Carmen Lucas-Estañ[1], Jan García-Morales[2], and Javier Gozalvez[1]
[1]*Universidad Miguel Hernández de Elche (UMH)*, Elche, Spain, [2]*Universidad Rey Juan Carlos (URJC)*, Madrid, Spain
Email: m.lucas@umh.es, jan.garcia@urjc.es, j.gozalvez@umh.es

***Abstract*—5G and beyond networks will support the digitalization of smart manufacturing thanks to their capacity to simultaneously serve different types of traffic with distinct QoS requirements. This can be achieved using Network Slicing that creates different logical network partitions (or slices) over a common infrastructure, and each can be tailored to support a particular type of traffic. The configuration of the Radio Access Network (RAN) slices strongly impacts the capacity of 5G and beyond to support critical services with stringent QoS requirements, and in particular deterministic requirements. Existing RAN Slicing solutions only consider the transmission rate (or bandwidth) requirements of the different services to partition the radio resources. This study demonstrates that this approach is not suitable to guarantee the stringent latency requirements of deterministic aperiodic traffic that is characteristic of industrial critical applications. We then propose designing RAN slices using descriptors that consider both the services' transmission rate and latency requirements, and demonstrate that this approach can support critical services that generate deterministic aperiodic traffic.**

## I. INTRODUCTION

Industry 4.0 leverages the integration of smart digital manufacturing cyber-physical systems and production systems for more intelligent manufacturing processes with zero defects. Industry 4.0 requires underlying communication networks capable to support a mix of services with different requirements (including high data rates, low latency and reliable connections) while providing the flexibility to support reconfigurable and adaptable production plants. 5G and beyond are considered key technical enablers for Industry 4.0 given their flexibility and capability to simultaneously support multiple services using a variety of technologies at the RAN and core networks.

The 5G Alliance for Connected Industries and Automation (5G-ACIA) and the 3GPP have identified and classified the Industry 4.0 use cases that can be effectively supported by 5G [1]. Industry 4.0 use cases present distinct communication requirements that can be categorized into three traffic classes: deterministic periodic, deterministic aperiodic, and non-deterministic [1]. Deterministic aperiodic traffic can be the most challenging traffic to support if it is time-critical and has stringent reliability requirements [1]. Its efficient management is a challenge in resource-constrained networks that have to simultaneously support multiple services [2]. Deterministic aperiodic traffic can occur sporadically at uncertain times and must be transmitted before a predetermined deadline. This class of traffic is associated with event-driven use cases where, for example, an event is activated when sensors detect malfunctions in devices, or a certain safety risk is detected in a production module [1]. In these scenarios, deterministic aperiodic traffic also demands low latency levels.

5G introduces significant innovations that help support mixed traffic services with different requirements. This includes a flexible 5G New Radio (NR) interface with different numerologies and Network Slicing (NS). NS exploits the softwarization and virtualization of networks to create different logical partitions or slices of a network over a common physical infrastructure. Each slice comprises a set of tailored network functions, radio access settings and resources to meet specific application Quality of Service (QoS) requirements. NS is particularly useful to simultaneously support multiple services with distinct QoS requirements [3], [4]. Network slicing can be implemented in the Core Network (CN) and in the RAN. RAN Slicing is particularly critical for applications that demand deterministic latency levels as the RAN significantly contributes to the end-to-end latency [5]. RAN slicing consists of the following phases: preparation, commissioning, operation, and decommissioning [6]. During the preparation phase, the application QoS requirements must be carefully assessed, and the slices are designed to support these requirements. In the commissioning phase, radio resources are allocated to the slices. The performance experienced is monitored during the operation phase, and the RAN slices might be adapted to continuously guarantee the application requirements. When the slices are no longer necessary, the slices are terminated.

An adequate design of the RAN slices is critical for meeting the latency requirements of deterministic traffic. However, so far, most proposals design RAN slices solely as a function of the transmission rate or bandwidth requirements of the services to be supported [7]-[10]. This study shows that this approach falls short in meeting the critical latency requirements of services that generate deterministic aperiodic traffic due to the sporadic or uncertain generation of packets. The study proposes designing RAN slices for deterministic aperiodic traffic using slice descriptors that account for both the transmission rate or bandwidth, and latency requirements. The study demonstrates that this approach can better support deterministic aperiodic traffic that is characteristic in critical vertical sectors, including Industry 4.0.

## II. Related Work

Most existing studies define RAN slices in terms of the number of radio resources necessary to support the transmission rate or bandwidth requirements of the services and nodes to be supported [7]-[10]. This is for example the case of the study in [7] that considers Guaranteed Bit Rate (GBR) services, and computes the resources necessary per slice based on aggregate rate requirements. The work in [8] also calculates the number of resources required by the RAN slices based on the GBR required by the services. [8] formulates the design of RAN slices as multiple ordinal potential games to minimize the average probability of blocking data sessions. The work in [9] combines resource-oriented (e.g. radio resource occupation levels) and rate-oriented (e.g. aggregated transmission rates) parameters to quantify the number of resources per slice. The proposal reported in [10] defines different RAN slices for best-effort, constant bit rate, and minimum bit rate services. The number of resources needed for each slice is calculated based on the average, constant and minimum transmission rate requirements of the different services. The authors predict the behavior of the cellular network using neural networks and adjust, accordingly, the RAN slices [10].

Previous proposals do not include latency requirements in the design of the slices, and hence cannot provide any latency guarantee. The proposal in [11] addresses this shortcoming with a novel latency-sensitive slice descriptor that identifies the number of radio resources per slice and their timing (in terms of TTIs, Transmission Time Intervals) in order to satisfy latency requirements for periodic traffic. The study in [12] uses this novel descriptor to propose a RAN slicing commissioning solution based on utility functions. In [13], authors include the latency requirement as a constraint of the optimization problem designed to dynamically allocate resources to RAN slices in a multi-tenant environment. In [14], authors take into account the latency requirements when computing the transmission rate that the RAN slice must guarantee. This rate is then used to estimate the number of radio resources that should be allocated to the slice. However, [14] does not establish any additional constraint about the timing (or TTIs) at which the radio resources should be reserved in order to satisfy the latency requirements. Existing latency-sensitive RAN slicing solutions clearly represent a step-forward, but do not still address deterministic latency constraints for aperiodic traffic characteristic of vertical sectors, such as Industry 4.0.

## III. Analytical Framework and Analysis

This section presents the analytical definition of the size and shape descriptors for the proposed latency-sensitive RAN slicing approach to support deterministic aperiodic traffic. The section also analytically derives the number of slices created in the slicing commissioning phase that can satisfy the latency requirements of deterministic aperiodic traffic with our proposed approach. We also compare this number with that obtained with conventional RAN slicing solutions that create slices only based on the size descriptor (i.e., the transmission rate requirements).

### A. Slice descriptors for deterministic aperiodic traffic

We consider that a RAN slice must support a group of $M$ nodes that generate deterministic aperiodic traffic in industrial environments. The nodes sporadically generate packets of small size $b$ that must be delivered before a latency deadline $D$ with reliability $P_{rel}$. Packets can be generated at any time, and the reliability is defined as the percentage of packets successfully delivered within the latency deadline required by the application.

The size of a slice for deterministic aperiodic traffic is given by the number of RBs needed to satisfy the transmission rate required by all the nodes. Each node $u$ needs to transmit a packet of size $b$ before a deadline $D$. The transmission rate needed for the transmission of each packet is then given by $R=b/D$. The effective transmission rate or throughput that can be achieved per each RB allocated to a node is given by $R_{eff} = TBS(SINR_u)/D$, where $SINR_u$ is the signal to interference plus noise ratio (SINR) experienced by node $u$, and $TBS(SINR_u)$ is the transport block size (TBS) or amount of data that can be transmitted per RB based on the experienced SINR. We should note that the TBS depends on the Modulation and Coding Scheme (MCS) utilized. We select the MCS based on the experienced SINR in order to guarantee a target BLER. The number $J_u$ of radio resources or RBs needed by each node $u$ can then be estimated as:

$$J_u = \frac{b/D}{TBS\,(SINR_u)/D} = \frac{b}{TBS\,(SINR_u)} \tag{1}$$

Deterministic aperiodic traffic in typical industrial applications is characterized by packet sizes between 20 and 40 bytes [1]. Considering the capabilities of 5G, we consider without loss of generality that these packets can be transmitted using 1 RB, i.e., $J_u$=1. $J_u$ radio resources should be reserved in every time window of size $D$ (expressed in number of slots) for a node $u$ to be able to meet the latency requirement for a packet that can be generated at any time instant. If the RAN slice must serve $M$ nodes, the allocation of $M{\cdot}J_u$ RBs to the slice in every time window of size $D$ would ensure meeting the latency requirement of all $M$ nodes regardless of the time instant at which the packets were generated. However, this approach overdimensions the slices and might result in an inefficient use and waste of the radio resources since packets are generated sporadically. To address this inefficiency, the 5G NR standard establishes the possibility of using semi-static scheduling [1] using shared resources [15]. In this case, radio resources or RBs are allocated to a group of nodes, and nodes must contend to gain access to the RBs for each packet transmission. This contention-based access can result in packet collisions. However, [16] demonstrated that the probability of collision can be reduced if the RBs are randomly selected among the set of available RBs in the time window between the generation time of a packet and the latency deadline $D$. Following [16], we consider that a node randomly selects the

[1] Configured Grant in Uplink (UL) or Semi-Persistent Scheduling (SPS) in Downlink (DL) allocate semi-static and periodic radio resources to nodes. In this case, nodes do not need to request radio resources for the transmission of each packet, which reduces the latency.

RB to transmit a packet among the $k$ available RBs. Packet generation for aperiodic traffic is typically modelled following a Poisson distribution with exponential inter-arrival time [16]; the average packet inter-arrival time is given by $\mu$. Following [16], the minimum amount of $k$ RBs that must be reserved in each time window of length $D$ to guarantee a probability of collision $P_c$ lower than 1-$P_{rel}$ is given by:

$$k = \frac{1\text{-}\exp(\text{-}D\cdot T_{slot}/\mu)}{1\text{-}P_{rel}^{\frac{1}{M\text{-}1}}} \quad (2)$$

where $T_{slot}$ is the slot duration in seconds, and it depends on the 5G NR numerology. The size $K$ of a slice can then be defined as:

$$K = \min\{k, M\cdot J_u\} \quad (3)$$

The shape of a slice determines how the RBs allocated to the slice must be distributed in time to meet the latency requirements of all the nodes supported by the slice. For deterministic aperiodic traffic, the slice shape must ensure that a packet generated at any time can be transmitted using $J_u$ RBs before its latency deadline $D$ and with a reliability $P_{rel}$. The slice shape identifies the slots over which the $K$ RBs allocated to the slice must be located. We define an allocation window of duration $W$ (expressed in number of slots) that establishes the time period during which the allocation of RBs to slices must be maintained [17]. We denote $L_t$ as the number of RBs allocated to the RAN slice in slot $t$. In this case, the slice shape for deterministic aperiodic traffic must satisfy the following latency condition:

$$\sum_{t=l}^{l+D\text{-}1} L_t = K,\ \ \forall l \in [1,W] \quad (4)$$

The expression in (4) establishes that $K$ RBs must be reserved within a time window of size $D$ starting at any slot $l$ of the allocation window.

### *B. Dimensioning of RAN slices*

The commissioning phase creates the RAN slices once they have been designed in the preparation phase. The preparation phase allocates the number of RBs necessary to each RAN slice, and the allocated RBs must comply with the slice descriptors (size and shape) quantified in the preparation phase. Depending on the number of available RBs within an allocation window, it could be possible to create multiple RAN slices that comply with the slice descriptors. This section analytically derives the number of RAN slices that can be created with the proposed latency-sensitive RAN slicing approach for deterministic aperiodic traffic. For comparison purposes, we also derive this number for conventional RAN slicing approaches that design the RAN slices exclusively based only on the transmission rate (and not the latency) requirements of the nodes.

We consider an allocation window of $W$ slots and a bandwidth that is divided in $N_{RB}$ RBs. Without loss of generality, we consider that $W$ is a multiple of the latency deadline $D$ (i.e., $W/D \in \mathbb{N}^+$). We first derive the number $N_p$ of RAN slices that can be created that jointly meet the size and shape descriptors proposed in (3) and (4). When the number of RBs allocated to the RAN slice in each slot $L_1, L_2, \ldots, L_W$ are known, the number of slices that can be created is $\prod_{t=1}^{W}\binom{N_{RB}}{L_t}$, where $\binom{N_{RB}}{L_t}$ represents the number of combinations of $L_t$ RBs within a resource grid including $N_{RB}$ RBs. Now, we calculate $N_P$ for any value of $L_1, L_2, \ldots, L_W$ that satisfies the size and shape descriptors. To this end, we first analyze the case where $D$=1. In this case, (4) establishes that $K$ RBs must be reserved in each slot $t$ of the allocation window, i.e. $L_t=K$, $\forall\, t \in [1,W]$. The number of RAN slices that satisfy the size and shape descriptors can then be calculated as:

$$N_P = \binom{N_{RB}}{K}^{W} \quad (5)$$

When $D$=2, the shape descriptor in (4) establishes that $L_1+L_2=K$, $L_2+L_3=K$, and $L_t+L_{t+1}=K$, $\forall t \in [1,W\text{-}1]$. From (4), we can establish that $L_{1+2i}=L_1\ \forall i \in [1,W/2\text{-}1]$, and $L_{2+2i}=K\text{-}L_1\ \forall i \in [1,W/2\text{-}1]$. In this case, $L_1$ can take any value between 0 and $\min\{K, N_{RB}\}$, and the number of RAN slices that satisfy the size and shape of the slice when $D = 2$ is given by:

$$N_P = \sum_{L_1=0}^{\min(K,\ N_{RB})} \binom{N_{RB}}{L_1}^{W/2} \binom{N_{RB}}{K\text{-}L_1}^{W/2} \quad (6)$$

We analyze in detail a last case with $D$=3 before deriving the general expression of $N_p$ for any value of $D$. When $D$=3, (4) establishes that $L_t+L_{t+1}+L_{t+2}=K\ \forall t \in [1,W\text{-}2]$. This results in $L_{1+3i}=L_1\ \forall i \in [1,W/3\text{-}1]$, $L_{2+3i}=L_2\ \forall i \in [1,W/3\text{-}1]$, and $L_{3+3i}=K\text{-}L_1\text{-}L_2\ \forall i \in [1,W/3\text{-}1]$. The number $N_p$ of RAN slices that can be created considering the size and shape descriptors when $D = 3$ is equal to:

$$N_P = \sum_{L_1=0}^{f(K)} \sum_{L_2=0}^{f(K\text{-}L_1)} \left[\binom{N_{RB}}{L_1}\binom{N_{RB}}{L_2}\binom{N_{RB}}{K\text{-}L_1\text{-}L_2}\right]^{W/3} \quad (7)$$

where function $f(x)=\min(x, N_{RB})$.

Based on (5), (6) and (7), the number $N_p$ of RAN slices that can be created following the size and shape descriptors in (3) and (4) when $D$ takes any value between 1 and $W$ can be derived as:

$$N_P = \sum_{L_1=0}^{f(K)} \cdots \sum_{L_{D\text{-}1}=0}^{f(K\text{-}\sum_{i=1}^{D\text{-}2} L_i)} \left(\prod_{j=1}^{D\text{-}1}\binom{N_{RB}}{L_j}\right)^{W/D} \cdot \binom{N_{RB}}{K\text{-}\sum_{j=1}^{D\text{-}1} L_j}^{W/D} \quad (8)$$

Most of the RAN slicing proposals in the literature (e.g. [8],[10]) design and create the RAN slices only considering the transmission rate requirements (i.e., only based on the size descriptor). The size of a RAN slice for deterministic aperiodic traffic is calculated in (3) as the number of RBs needed to guarantee the transmission rate requirements within a time window of size equal to the latency requirement $D$. In this case, the number of RBs that must be allocated to the RAN slice in an allocation window of size $W$ is equal to $K\cdot W/D$ when $W=n\cdot D$ and $n \in \mathbb{N}^+$. None of the reference size-based RAN slicing proposals ([8],[10]) address the case of deterministic aperiodic traffic and establish constraints regarding the timeslots where the RBs must be reserved inside the alçlocation window. We then consider two options for the implementation of the RAN slicing approach that designs slices exclusively based on the size descriptor:

1. *Size1*: we consider that there are no constraints and the $K \cdot W/D$ RBs can be reserved in any of the slots within the allocation window $W$. In this case, the number $N_{Size1}$ of RAN slices that can be created is:

$$N_{Size1} = \binom{N_{RB} W}{K(W/D)} \quad (9)$$

2. *Size2*: we consider that $K$ RBs from the total $K \cdot W/D$ RBs that are allocated to a RAN slice must be reserved in each period of $D$ slots within the allocation window $W$, i.e., $K = \sum_{t=1+i \cdot D}^{i \cdot D+D} L_t$, $\forall\, i=0, \ldots, W/D$. This criterion is applied in [11] for periodic traffic, and we aim to evaluate its capability to support aperiodic traffic. In this case, the number $N_{Size2}$ of RAN slices that can be created following this criterion is given by:

$$N_{Size2} = \prod_{p=1}^{W/D} \binom{N_{RB} D}{K} = \binom{N_{RB} D}{K}^{W/D} \quad (10)$$

From (8), (9) and (10), we can see that $N_P \leq N_{Size2} \leq N_{Size1}$. In fact, the set of RAN slices that can be created following *Size2* is a subset of the set of RAN slices that can be created following *Size1*. Furthermore, the set of RAN slices created considering both the size and shape descriptors is a subset of the set of RAN slices created following *Size2*.

### *C. Capacity to satisfy the latency requirements*

Fig. 1 shows the percentage of RAN slices that satisfy the latency requirement for deterministic aperiodic traffic in a scenario with an allocation window of 10 slots, $N_{RB}$=25 RBs, and the latency deadline $D$ equal to 2 or 5 slots. The figure compares the percentage achieved with our proposed latency-sensitive RAN slicing approach (*Size&Shape*) and with the approaches that design RAN slices solely based on transmission rate requirements (*Size1* and *Size2*). The percentage is depicted as a function of the size $K$ of the slice. Fig. 1 shows that all the RAN slices that can be created when jointly considering the size and shape descriptors ($N_p$ in (8)) meet the latency requirement expressed in (4). On the other hand, the percentage decreases with $K$ for the RAN slicing schemes that create slices considering only the transmission rate requirements. In general, *Size2* achieves better performance than *Size1*. However, Fig. 1 clearly shows that considering the size and shape descriptors to design RAN slices significantly improves the capability to satisfy latency requirements compared to only considering the size descriptor (independently of the implementation option). Fig. 1 also shows that the percentage of RAN slices that satisfy the latency requirements decreases with $D$ when RAN slices are created only considering the size descriptor. This is not the case when $D$=1 and RAN slices are created following the *Size2* implementation. In this case, *Size2* allocates $K$ RBs in each slot within the allocation window in a similar way as when the RAN slices are created considering both the size and shape descriptors. This results in that all RAN slices created using *Size2* can meet the latency requirement. In fact, it can be observed from (5) and (10) that $N_p$=$N_{Size2}$ when $D$=1.

Fig. 2 depicts the percentage of RAN slices that satisfy the latency requirement for deterministic aperiodic traffic as a function of the size $W$ of the allocation window. The figure shows that all the RAN slices created using the size and shape descriptors (*Size&Shape*) satisfy the latency requirements for any value of $W$. When $W$=$D$, $N_p$=$N_{Size2}$=$N_{Size1}$ and the three RAN slicing approaches (*Size&Shape, Size1* and *Size2*) satisfy the latency requirement. However, Fig. 2 shows that this is not the case for other values of $W$ as RAN slices created using only the size descriptor cannot satisfy the latency requirement for all the configurations reported under Fig. 2.

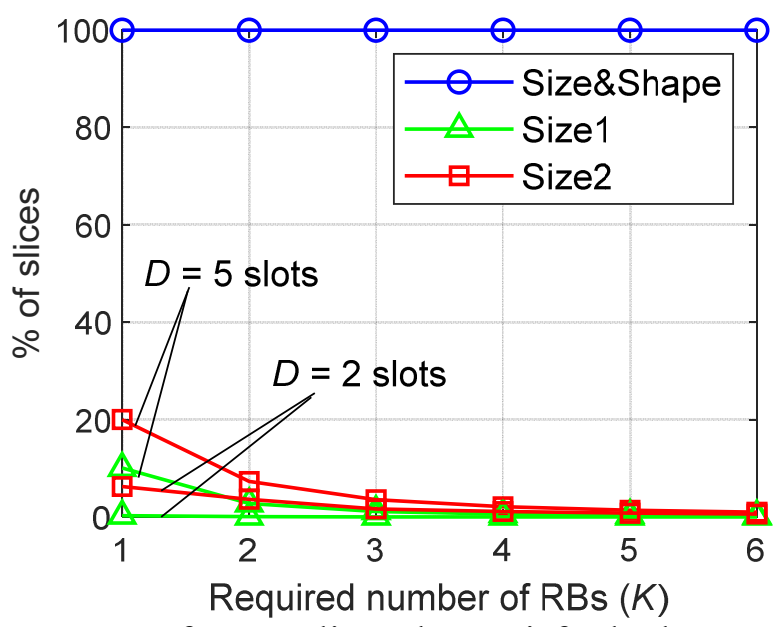

Fig. 1. Percentage of RAN slices that satisfy the latency requirement as a function of the size $K$ of the slice ($N_{RB}$=25 RBs and $W$=10 slots).

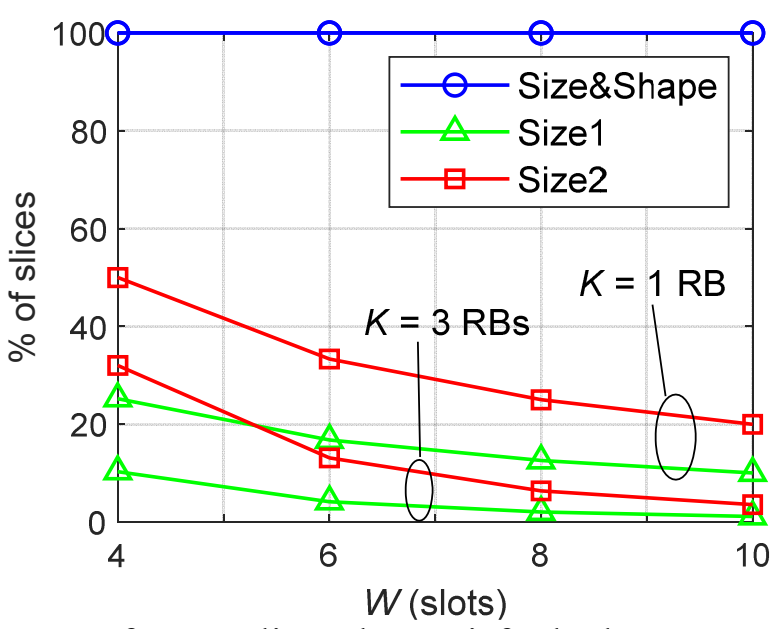

Fig. 2 Percentage of RAN slices that satisfy the latency requirement as a function of the size $W$ of the allocation window ($N_{RB}$=25 RBs, $D$=2 slots).

## IV. Simulation Evaluation

The previous section numerically analyzed the capacity of RAN slices to satisfy the latency requirement when slices are designed considering the size and shape descriptors or only the size descriptor. The analysis has shown that only a low percentage of the RAN slices created using the size descriptor can guarantee the latency requirement established in (4). While these slices may not satisfy the latency requirement for all the nodes it should support, they could still fulfill the latency requirement for some of the nodes. This section complements the previous numerical analysis with a Monte-Carlo simulation study that quantifies the percentage of transmissions for which each RAN slicing scheme under evaluation can satisfy the latency requirements. We emulate an industrial plant with $M$=20 nodes uniformly distributed that generate deterministic aperiodic traffic. Each node transmits packets with a size of 20 bytes [12], and packets are generated following a Poisson distribution with exponential inter-arrival time equal to 5 seconds ($\mu$=5) [16]. A packet transmission is considered successful if it receives $J_u$ RBs and the transmission is completed before the latency deadline $D$. We consider a 5G Non-Public-Network (NPN) with a single cell that covers the industrial plant. Data is transmitted in UL using Configured Grant. With Configured Grant, radio resources are pre-allocated to the nodes when the session is established, and nodes can transmit their packets as soon as the data is generated [2][18]. 5G NR is configured with a subcarrier spacing of 30 kHz [18] (the slot duration $T_{slot}$ is equal to 0.5 ms), and a bandwidth of 40 MHz that is divided in $N_{RB}$= 106 RBs. We consider a target BLER equal to $10^{-5}$ following [15].

Fig. 3 depicts the average percentage of successful packet transmissions as a function of *D*. Fig. 3 shows that creating RAN slices using the size and shape descriptors (*Size&Shape*) satisfies the latency and transmission rate requirements for all the transmissions. The capacity to satisfy both requirements when the RAN slices are created only based on the size descriptor strongly depends on the implementation option followed. With *Size1*, Fig. 3 shows that the percentage of successful transmissions is in general quite low but increases with *D*. On the other hand, *Size2* guarantees a very high percentage of successful transmissions for all values of *D*. In fact, it can achieve similar satisfaction levels to *Size&Shape*. We should note that the percentage of successful transmissions is significantly higher for *Size1* and *Size2* than the results depicted in Section III.C. This is because even if a slice does not meet the latency requirement for all the nodes, it can still successfully support a relatively large percentage of transmissions. However, Fig. 3 confirms the conclusion observed in Section III.C, since the only option to always satisfy both the transmission rate and latency requirements is to consider the size and latency descriptors proposed by the authors when creating the RAN slices.

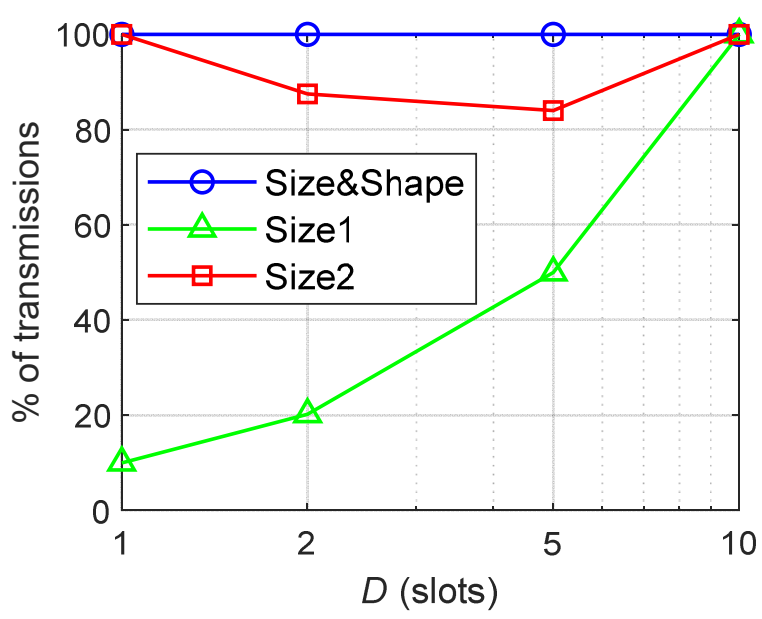

Fig. 3 Average percentage of successful transmissions as a function of *D* (*W*=10 slots).

## V. Conclusions

This study has analyzed the possibility to design 5G-based RAN slices capable to satisfy the transmission rate and latency requirements of applications that generate deterministic aperiodic traffic. Supporting such traffic is important for critical vertical scenarios such as Industry 4.0 or smart manufacturing. The study has demonstrated that creating RAN slices based only on transmission rate requirements cannot guarantee the latency requirements for all transmissions in critical applications with deterministic aperiodic traffic. This is though possible with the authors' proposal to design RAN slices using the size and shape descriptors. Future work includes the proposal of new RAN slice descriptors to account for reliability requirements, and the study of solutions to create RAN slices simultaneously accounting for transmission rate, latency, and reliability requirements.

## Acknowledgments

This work has been funded by European Union's Horizon Europe Research and Innovation programme under the Zero-SWARM project (No 101057083), by MCIN/AEI/10.13039/501100011033 through the project PID2020-115576RB-I00, and by Generalitat Valenciana (CIGE/2022/17), and UMH's Vicerrectorado de Investigación grants (VIPROAS23/11).